\documentclass[
    aps,prc,
    floatfix,
    reprint,
    amsfonts,amssymb,amsmath,
    superscriptaddress,
]{revtex4-2}

\usepackage{orcidlink}
\usepackage[all]{hypcap}
\usepackage{graphicx}
\usepackage{physics}
\usepackage{enumitem}
\hypersetup{
    colorlinks=true,
    linkcolor=blue,
    urlcolor=magenta,
    citecolor=blue
}

\newcommand{\RAA}{\ensuremath{R_{AA}}}
\newcommand{\sNN}{\ensuremath{\sqrt{s_{NN}}}}
\newcommand{\pTjet}{\ensuremath{p_\mathrm{T}^\mathrm{jet}}}
\newcommand{\rg}{\ensuremath{r_{g}}}
\newcommand{\pp}{\ensuremath{{pp}}}
\renewcommand{\AA}{\ensuremath{{AA}}}
\newcommand{\PbPb}{\ensuremath{{\rm Pb+Pb}}}
\newcommand{\vbp}[1][]{\ensuremath{\Vec{b}_{#1 \perp}}}
\newcommand{\vrp}[1][]{\ensuremath{\Vec{r}_{#1 \perp}}}
\newcommand{\pt}[1][]{\ensuremath{p_\mathrm{T}^{ #1}}}
\newcommand{\qt}{\ensuremath{q_\mathrm{T}}} 
\newcommand{\lt}{\ensuremath{l_\mathrm{T}}}
\newcommand{\kt}{\ensuremath{k_\mathrm{T}}}
\newcommand{\vpt}[1][]{\ensuremath{\Vec{p}_\mathrm{T}^{ #1}}}
\newcommand{\vlt}{\ensuremath{\Vec{l}_\mathrm{T}}}
\newcommand{\vqt}{\ensuremath{\Vec{q}_\mathrm{T}}}
\newcommand{\vkt}{\ensuremath{\Vec{k}_\mathrm{T}}}
\newcommand{\dij}{\ensuremath{d_{ij}}}
\newcommand{\dc}{\ensuremath{d_{\rm c}}}
\newcommand{\thetaij}{\ensuremath{\theta_{ij}}}
\newcommand{\thetac}{\ensuremath{\theta_{\rm c}}}
\newcommand{\Qij}{\ensuremath{Q_{ij}}}
\newcommand{\Qc}{\ensuremath{Q_\mathrm{c}}}
\newcommand{\Qmed}{\ensuremath{Q_\mathrm{med}}}
\newcommand{\Ico}{\ensuremath{I_\mathrm{co}}}
\newcommand{\Rjet}{\ensuremath{R_\mathrm{jet}}}
\newcommand{\Rsubjet}{\ensuremath{R_\mathrm{subjet}}}
\newcommand{\as}{\ensuremath{\alpha_\mathrm{s}}}
\newcommand{\zcut}{\ensuremath{z_\mathrm{cut}}}

\begin{document}

\title{A resolution-based framework for coherent and decoherent jet energy loss in NLO pQCD}

\author{Peng-Hui \surname{Hu} \orcidlink{0009-0009-9328-0665}}
\affiliation{Key Laboratory of Quark and Lepton Physics (MOE) and Institute of Particle Physics, Central China Normal University, Wuhan 430079, China}

\author{Zhan \surname{Gao} \orcidlink{0000-0002-3194-4886}}
\affiliation{Three Gorges Tourism Polytechnic College, Yichang, Hubei 443111, China}

\author{Lin \surname{Chen} \orcidlink{0000-0003-2082-3533}}
\email{zhanghz@mail.ccnu.edu.cn}
\email{lin.chen@usc.es}
\affiliation{Instituto Galego de F\'isica de Altas Enerx\'ias IGFAE, Universidade de Santiago de Compostela, E-15782 Galicia, Spain}

\author{Liqiang \surname{Zhu} \orcidlink{0009-0004-3826-9395}}
\affiliation{Key Laboratory of Quark and Lepton Physics (MOE) and Institute of Particle Physics, Central China Normal University, Wuhan 430079, China}
\affiliation{State Key Laboratory of Nuclear Physics and Technology, Institute of Quantum Matter, South China Normal University, Guangzhou 510006, China}
\affiliation{Guangdong Basic Research Center of Excellence for Structure and Fundamental Interactions of Matter, Guangdong Provincial Key Laboratory of Nuclear Science, Guangzhou 510006, China}

\author{Hanzhong \surname{Zhang} \orcidlink{0000-0002-4884-4531}}
\affiliation{Key Laboratory of Quark and Lepton Physics (MOE) and Institute of Particle Physics, Central China Normal University, Wuhan 430079, China}

\begin{abstract}
We investigate the role of medium resolution and color decoherence in jet energy loss in high-energy heavy-ion collisions.
Within an NLO pQCD framework, implemented using \texttt{NLOJet++}, we calculate inclusive jet production in $\pp$ collisions and extend the calculation to $\AA$ collisions by incorporating nuclear PDFs, Glauber geometry, and medium-induced jet energy loss in the Higher-Twist formalism.
To account for the finite resolution of the QGP, we introduce a sequential coherent clustering procedure based on the pairwise invariant mass of final-state partons.
The resulting coherent clusters act as independent color sources when resolved by the medium, with the transition governed by an effective resolution scale $\Qmed$.
We validate the coherent clustering procedure using \texttt{PYTHIA} simulations and investigate the nuclear modification factors of inclusive jets, reclustered jets, soft-drop $\rg$, and jet mass in $\PbPb$ collisions at $\sNN=5.02$ TeV.
Comparison with the ATLAS measurements yields $\Qmed=10-20$ GeV.
While coherent energy loss alone systematically underestimates the observed suppression at high jet $\pTjet$ and fails to reproduce the substructure dependence of jet quenching, incorporating decoherent energy loss provides a consistent description of these observables.
Our results demonstrate that accounting for the medium's ability to resolve jet substructure is essential for a consistent description of jet quenching in heavy-ion collisions at the LHC.
\end{abstract}

\maketitle


\section{Introduction}
\label{sec:intro}

The Relativistic Heavy-Ion Collider (RHIC) at Brookhaven National Laboratory (BNL) and the Large Hadron Collider (LHC) at European Organization for Nuclear Research (CERN) can accelerate two beams of heavy-ions to collide at almost the speed of light.
Partons become deconfined during high-energy heavy-ion collisions, forming a hot dense medium known as the Quark-Gluon Plasma (QGP).
In these collisions, partons from hard scatterings with large momentum fragment into collimated sprays of particles, known as jets. 
Energetic jets traverse the QGP medium and interact with its thermal partons, resulting in their energy loss and transverse momentum broadening, a phenomenon called jet quenching~\cite{Gyulassy:1990ye, Wang:1992qdg, Wiedemann:2009sh, Armesto:2011ht, Qin:2015srf, Cao:2024pxc}.

Compared with proton+proton ($\pp$) collisions, the jets produced in nucleus+nucleus ($\AA$) collisions exhibit yield suppression and modifications to their substructure.
The influence of jet quenching can be directly observed through certain observables, such as the nuclear modification factor $\RAA$~\cite{ATLAS:2014ipv}, jet shape~\cite{Luo:2021hoo}, soft-drop $\rg$~\cite{Larkoski:2014wba}, jet mass~\cite{Gao:2025fch,Duan:2023gmp}, and energy-energy correlators (EEC)~\cite{Yang:2026bsk}.

Various theoretical models have been developed to quantify the jet energy loss in the QGP.
These include the BDMPS-Z formalism~\cite{Zakharov:1996fv, Baier:1998kq, Baier:1996kr, Zakharov:1997uu}, GLV~\cite{Gyulassy:2000fs, Gyulassy:2000er} and higher-twist (HT) approach~\cite{Guo:2000nz, Wang:2001ifa, Zhang:2003yn, Deng:2009ncl, Wang:2016fds}.
Jet transport models based on Monte Carlo approaches have also been developed, such as AMPT~\cite{Lin:2004en, Zhang:2005ni, Ma:2010dv,Luo:2024xog,Duan:2025wsy}, LBT~\cite{He:2015pra, Cao:2016gvr, Cao:2017hhk, Luo:2018pto, Dang:2026ezw}.
The strength of jet quenching is commonly characterized by the jet transport coefficient $\Hat{q}$~\cite{JET:2013cls, Qin:2015srf, Xie:2022ght, Xie:2022fak}, which is defined as the mean transverse momentum squared exchanged between the energetic parton and medium per unit length.
$\Hat{q}$ is sensitive to the local properties of the QCD medium, such as the temperature and the flow velocity.

With the recent upgrade of experimental facilities, the nucleon–nucleon center-of-mass energy $\sNN$ in LHC heavy-ion collisions has reached $5.02$ TeV, and measured jet transverse momenta $\pTjet$ have extended up to 1000 GeV~\cite{ATLAS:2018gwx, ATLAS:2023hso, ATLAS:2022vii, ATLAS:2018jsv}. 
In particular, new measurements including reclustered jets~\cite{ATLAS:2023hso}, soft-drop $\rg$~\cite{ATLAS:2022vii}, and jet mass~\cite{ATLAS:2018jsv} can provide new experimental resources for investigating the impact of jet quenching on jet substructure. 
These energy scales far exceed those of previous experiments, offering a robust foundation for probing new jet-medium regimes.

In traditional theoretical approaches~\cite{Wang:2016fds}, the jet is assumed to lose energy as a single coherent entity, due to the fact that the medium cannot resolve the substructures of the jet, which was a valid assumption at lower energies.
However, a jet may contain multiple coherent clusters in high-energies that lose energy independently~\cite{Mehtar-Tani:2010ebp, Casalderrey-Solana:2011ule, Mehtar-Tani:2012mfa, Mehtar-Tani:2017web}, and this decoherent effect becomes important in jet energy loss calculations in this regime.
Color coherence and decoherence effects have been investigated by researchers employing various approaches~\cite{Casalderrey-Solana:2012evi, Kudinoor:2025ilx, Cunqueiro:2023vxl, Duan:2026nvr,Caucal:2018dla,Andres:2023xwr,Mehtar-Tani:2021fud,Caucal:2026lvn,Pablos:2024muu,Casalderrey-Solana:2019ubu,Dominguez:2019ges,Kuzmin:2025fyu,Andres:2026tas,Lamas:2026ntq,Lamas:2026ntq,Dokshitzer:1987nm,Mehtar-Tani:2011hma,Mehtar-Tani:2011vlz}, such as characterizing the jet-medium resolution scale by using the critical angle $\thetac$ as the criterion for decoherence~\cite{Casalderrey-Solana:2012evi,Abreu:2024wka}, or calculating the number of subjets generated via vacuum-like emissions up to scale $Q_0$~\cite{Duan:2026nvr,Duan:2025lvi}.
Furthermore, additional approaches, such as resummation~\cite{Mehtar-Tani:2017web}, EFT~\cite{Vaidya:2026yfa}, JEWEL~\cite{Zapp:2026cqf}, and Hybrid~\cite{Hulcher:2017cpt}, have been employed, along with the application of observables like EEC~\cite{Andres:2022ovj,Barata:2023bhh,Ke:2025ibt,Bossi:2026mdz,Mehtar-Tani:2026bmr}, Groomed jets~\cite{Mehtar-Tani:2016aco,Caucal:2019uvr,Caucal:2021cfb}, and subjets~\cite{Mehtar-Tani:2017ypq}.

In this work, we introduce a coherent clustering procedure for the identification of coherent clusters from final-state partons.
This clustering algorithm is based on the resolution parameter of the medium called $\Qmed$ which quantifies the coherent clusters that the medium can resolve.
These coherent clusters then lose energy independently and modifies the measured inclusive and reclustered jet $\RAA$, and jet substructure observables.

The paper is organized as follows:
In~\autoref{sec:theory}, the differential cross section of $\pp$ and $\AA$ collisions at NLO pQCD will be represented.
Then, jet energy loss is described under HT formalism.
Most importantly, for coherent and decoherent jet quenching, we will introduce the coherent clustering procedure, the resolution scale $\dij$, and the coherent intensity $\Ico(\Qij)$ for the $\Qij$ criterion.
In~\autoref{sec:result}, we first generate coherent clusters in $\pp$ collisions using \texttt{PYTHIA8} simulation to validate our approach.
To investigate the $\pTjet$ dependence of jet suppression, we calculate the $\RAA$ of inclusive jets and reclustered jets as a function of $\pTjet$ in $\PbPb$ collisions at $\sNN=5.02$ TeV.
To explore the jet substructure dependence of jet suppression, we further calculate the $\RAA$ as a function of soft-drop $\rg$ and jet mass in the same collisions system.
The theoretical predictions for these observables are compared with experimental measurements, for which we extracted $\Qmed = 10-20$ GeV.
Finally, the conclusions are presented in~\autoref{sec:conclusion}.


\section{Theoretical setup}
\label{sec:theory}


\subsection{Cross section and measurement function}
\label{subsec:xsec}

In collinear factorization, the single-inclusive jet production cross section in proton+proton ($\pp$) collisions can be expressed as a convolution of parton distribution functions (PDFs) with the short-distance partonic cross section as
\begin{align}
\label{eq:pp-xsec}
\frac{\dd\sigma_{\pp\to{\rm jet}}}{\dd\pTjet\dd y}
&=\sum_{a,b}\int\dd x_a\dd x_b~\frac{\dd\Hat{\sigma}_{ab\to{\rm jet}}}{\dd\pTjet\dd y}\notag\\
&\times f_{a/p}(x_a,\mu_F)f_{b/p}(x_b,\mu_F).
\end{align}
The collinear PDFs $f_{i/p}(x_i,\mu_F)$ are evaluated at momentum fraction $x_i$ and factorization scale $\mu_F$.
The partonic cross section $\dd\Hat{\sigma}$ contains both leading-order (LO) and next-to-leading-order (NLO) contributions in perturbative QCD (pQCD).
In~\autoref{sec:result}, the LO is defined as the $2\to2$ Born process for inclusive and reclustered jets, whereas for $\rg$ and jet mass, it is taken as the $2\to3$ Born process.

It is useful to write the partonic cross section in a form that makes explicit the dependence on the measurement function.
\begin{equation}
\dd\Hat{\sigma}_{2\to n}=\frac{1}{2\Hat{s}}\left|\mathcal{M}_{2\to n}\right|^2\dd\Phi_n~\mathcal{S}_n(\{p_i\},R),
\end{equation}
where the Mandelstam variable $\Hat{s}$ is the partonic hard scale, $|\mathcal{M}_{2\to n}|^2$ is the matrix element, and $\dd\Phi_n$ is the phase space measure.
For each final-state multiplicity $n$, the short-distance matrix element is integrated over the corresponding phase space weighted by a measurement operator $\mathcal{S}_n$ that defines the observable under consideration.
For single-inclusive jet production, this operator depends on the final-state momenta $\{p_i\}$ and on the jet clustering parameter $R$.
The measurement function encodes the jet algorithm and projects the partonic final-state onto the observed jet transverse momentum $\pTjet$ and rapidity $y$.

Writing the cross section in this operator form makes explicit that all information about how partons are mapped into jets is contained in the measurement function, while the hard matrix elements remain universal.
In the present calculation, the partonic matrix elements and phase-space integration are evaluated using the \texttt{NLOJet++} framework~\cite{Nagy:2003tz, NLOJet}, which implements the subtraction formalism required to ensure infrared and collinear finiteness.
Our framework then focuses on the definition and modification of the measurement function $\mathcal{S}_n$ at NLO.

The measurement function for single-inclusive jet can be explicitly written.
For a two-parton final state, it is simply $\mathcal{S}_2=\sum_{i}\delta(\pTjet-\pt[i])$, due to the fact that the two partons are almost back-to-back.
The appropriate kinematics constraints and acceptance cuts are implicitly understood.
For three-particles, the measurement function is heuristically written as:
\begin{align}\label{eq:pp-Sn}
\mathcal{S}_3
&=\sum_{i,j,k} \frac{s_{ijk}}{2}\left[\Theta_{ij}\delta(\pTjet-\pt[ij])+\Theta_{k}\delta(\pTjet-\pt[k])\right]
\end{align}
where $s_{ijk}$ denotes the symmetric tensor, $\Theta_{ij}(R)=1$ when partons $i$ and $j$ can be clustered into a jet with $\pt[ij]=|\vpt[i]+\vpt[j]|$, and $\Theta_k(R)=1$ when parton $k$ is isolated.
The clustering condition $\Theta(R)$ is dependent on the jet algorithm applied, which depends on the angular separation $\Delta R_{ij}^2=(\phi_i-\phi_j)^2+(y_i-y_j)^2$ and the jet clustering parameter $R$.
It is important to note that the measurement function is infrared and collinear (IRC) safe, such that $\mathcal{S}_3\to\mathcal{S}_2$ when one of the particles is soft or collinear to another particle.
This property ensures the cancellation of singularities between real and virtual corrections and guarantees the consistency of the NLO calculation.

The single-inclusive jet cross section in $\AA$ collisions is obtained by extending the corresponding $\pp$ baseline to incorporate nuclear structure effects, medium-induced modifications to the measurement function, and the geometrical profile of the heavy-ion collision
\begin{align}
\label{eq:aa-xsec}
\frac{\dd\sigma_{AA\to{\rm jet}}}{\dd\pTjet\dd y}
&=\sum_{a,b}\int_{\mathcal{C}}\dd^2\vbp T_{AA}(\vbp)\int\frac{\dd\phi}{2\pi}\notag\\
&\times\int\dd x_a\dd x_b~f_{a/A}(x_a,\mu_{F})f_{b/A}(x_b,\mu_{F})\notag\\
&\times\sum_{n}\int\dd\Phi_n~\frac{1}{2\Hat{s}}\left|\mathcal{M}_{2\to n}\right|^2\Tilde{\mathcal{S}}_n(\{p_i\},R).
\end{align}

The first line in~\autoref{eq:aa-xsec} encodes the geometrical averaging over the transverse plane.
For a fixed centrality class $\mathcal{C}$, the hard scattering probability is integrated over the transverse production plane, weighted by the nuclear overlap function 
\begin{equation}
\label{eq:TAA}
T_{AA}(\vbp)=\int\dd^2\vrp T_A(\vrp+\vbp/2)T_A(\vrp-\vbp/2)   
\end{equation}
obtained within the Glauber model~\cite{Jacobs:2000wy}, with $T_{A}$ the thickness function.
Consequently, the medium modification experienced by the jet depends explicitly on both its production coordinate $\vrp$ and its propagation direction $\phi$, which together determine the in-medium path length and local density profile encountered by the parton.

The second line in~\autoref{eq:aa-xsec} implements nuclear structure effects through the replacement of the proton PDFs by nuclear PDFs (nPDFs)
\begin{align}
f_{a/A}(x,\mu_{F})
&=\frac{Z}{A}~R^{p/A}_a(x,\mu_{F})f_{a/p}(x,\mu_{F})\notag\\
&+\frac{A-Z}{A}~R^{n/A}_a(x,\mu_{F})f_{a/n}(x,\mu_{F}),
\end{align}
where the proton PDFs $f_{a/p}(x)$ are obtained from \texttt{CTEQ}~\cite{Pumplin:2002vw,Giele:2001mr} and the nuclear modification factor $R_a^{p/A}(x)$ is provided by \texttt{EPPS21}~\cite{Eskola:2021nhw}.
The corresponding neutron PDF and nuclear modification factor can be derived by assuming isospin symmetry.
This substitution ensures a consistent treatment of the initial-state nuclear effects at the factorization scale $\mu_F$.
It has been shown that the effects of nPDFs can significantly influence the final particle yield at high transverse momentum~\cite{Adhya:2021kws}.

The medium modified measurement function $\Tilde{\mathcal{S}}_n$ in $\AA$ collisions can be written similar to the vacuum case, but with parton transverse momentum replaced by the quenched momentum, i.e. $\pt[ij]\to\mathcal{Q}(\pt[ij])$ and $\pt[k]\to\mathcal{Q}(\pt[k])$, where $\mathcal{Q}(\pt)=\pt-\Delta E(\pt)$, and $\Delta E$ is the jet energy loss.
For jets with one parton, the parton is treated as a single coherent color charge with energy loss dependent on the parton's transverse momentum $\Delta E(\pt[i])$.
For two partons that are combined into a single jet, energy loss is dependent on the momentum sum of the two partons $\Delta E(\pt[ij])$.
In other words, we perform the coherent clustering first and apply energy loss to the transverse momentum of this coherent cluster.

It is important to emphasize that, within this framework, the hard matrix elements and phase-space structure remain identical to those in $\pp$ collisions.
All medium effects are encoded in the nPDFs and in the modified measurement function, which incorporates the deterministic energy loss shift of the jet momentum.
This separation preserves the perturbative consistency of the calculation while allowing for a transparent implementation of medium-induced modifications such that jet-medium interaction is assumed to be factorized from the hard scattering.


\subsection{Jet energy loss}
\label{subsec:eloss}

Energetic partons produced in initial hard scatterings propagate through the QGP created in high-energy heavy-ion collisions.
During their propagation, they undergo multiple soft interactions with the medium constituents.
These interactions induce transverse momentum broadening and medium-induced gluon radiation.
The resulting redistribution of energy and momentum modifies both inclusive hadron spectra and fully reconstructed jets, a phenomenon commonly referred to as jet quenching~\cite{Gyulassy:1990ye, Wang:1992qdg, Wiedemann:2009sh, Armesto:2011ht, Qin:2015srf, Cao:2024pxc}.
In perturbative description of in-medium propagation, the microscopic interaction strength between a fast parton and the QGP is encoded in a single transport coefficient, known as the jet quenching parameter~\cite{JET:2013cls}
\begin{equation}
\Hat{q}\equiv\frac{\dd \langle \qt^2\rangle}{\dd L},
\end{equation}
defined as the average squared transverse momentum $\langle \qt^2\rangle$ transferred from the medium per unit path length $L$.
Physically, $\Hat{q}$ characterizes the transverse color Lorentz force exerted by the fluctuating chromodynamic fields of the plasma.
It is sensitive to both the local temperature and the color representation of the propagating parton.
In general, partons lose energy through both elastic (collisional) and inelastic (radiative) processes. 
For highly energetic jets with $E\gg T$, medium-induced gluon radiation provides the dominant contribution to the total energy loss, while elastic processes primarily contribute to transverse diffusion and subleading energy degradation~\cite{Qin:2007rn}. 
In the following, we focus on radiative energy loss within the higher-twist (HT) framework, while consistently incorporating transverse momentum diffusion through $\Hat{q}$.

The HT formalism treats medium-induced gluon emission as a higher-twist correction to standard perturbative QCD factorization.
Physically, the medium modifies the parton splitting process through rescattering of the outgoing parton-gluon system with color fields of the plasma.
The resulting radiation spectrum can be interpreted as arising from the interference between vacuum emission and scattering-induced emission amplitudes.
Within this approach, the single-gluon emission spectrum from a parent parton $a=q,g$ is given as~\cite{Deng:2009ncl}
\begin{align}
\label{eq:ht-ker}
\frac{\dd N_a}{\dd z\dd\lt^2\dd\tau}
&=\frac{2C_A\as}{\pi\lt^4}\left[\frac{1+(1-z)^2}{z}\right]\notag\\
&\times \Hat{q}_a(\tau)\sin^2\left[\frac{\lt^2(\tau-\tau_0)}{4z(1-z)E}\right],
\end{align}
where $z$ and $\lt=|\vlt|$ are the energy fraction and transverse momentum carried by the emitted gluon respectively, and $E$ is the parent parton energy.
The local transport coefficient $\Hat{q}_a(\tau)$ indicates that the emission rate is proportional to the density of scattering centers at time $\tau$, embedding the space-time structure of the evolving QGP into the radiation kernel.
The sine squared factor encodes the Landau-Pomeranchuk-Migdal (LPM) interference effect~\cite{Wang:1994fx}, and suppresses emissions with formation time $\tau_f=2z(1-z)E/\lt^2$ larger than the local path length $\tau-\tau_0$.
Thus, radiation is dependent also on the accumulated propagation history.
The local transport coefficient is evaluated perturbatively from elastic scattering with thermal partons~\cite{He:2015pra}
\begin{equation}
\label{eq:qhat-local}
\Hat{q}_a(\tau)=
C_a\frac{42\zeta(3)}{\pi}\as^2T^3 \ln\left(\frac{s^*}{4\mu_D^2}\right),
\end{equation}
with $C_q=C_F$, $s^*=5.8ET$ for quark jet and $C_g=C_A$, $s^*=5.6ET$ for gluon jet.
$\zeta(3)\approx1.202$ is the Ap\'ery's constant.
$\mu_D$ is the Debye screening mass~\cite{He:2015pra}, $\mu_D^2 = (N_c/3 + N_f/6) g^2 T^2$, where $N_f=3$ is the number of active quark flavors in the QGP and $g^2=4\pi\as$.
It is important to note that the strong coupling $\as$ in $\mu_D$,~\autoref{eq:ht-ker}, and~\autoref{eq:qhat-local} are identical and extracted from experimental data.
The local temperature field $T(\tau,\vrp)$ is supplied by the CLVisc (3+1)D relativistic hydrodynamic model~\cite{Pang:2012he, Pang:2014ipa, Pang:2018zzo}, with the initial evolution time of the medium $\tau_0=0.6$ fm. 

In addition to the medium induced radiation, multiple soft scatterings lead to transverse momentum diffusion of the radiated gluon.
In the soft-scattering limit, where individual momentum transfers are small compared to the jet energy, the cumulative effect can be treated as a diffusion process in transverse momentum space.
The total accumulated transverse momentum squared is given by
\begin{equation}
\langle\qt^2\rangle_a=\int_{\tau_0}\dd\tau~\Hat{q}_a(\tau).
\end{equation}
This relation follows directly from the definition of $\Hat{q}$ and expresses the fact that broadening builds up additively along the path.
We model the probability distribution for the total transverse momentum kick $\qt=|\vqt|$ as a Gaussian form
\begin{equation}
F(\qt)=\frac{1}{2\pi\langle\qt^2\rangle}\exp\left(-\frac{\qt^2}{2\langle\qt^2\rangle}\right),
\end{equation}
normalized as $\int\dd^2\vqt~F(\qt)=1$.
This corresponds to the solution of a Fokker-Planck equation in transverse momentum space, which is a valid approximation when multiple independent soft scatterings dominate over rare hard kicks.
Consider now a gluon emitted with intrinsic transverse momentum $\vlt$ relative to the jet axis.
After subsequent diffusion, the gluon acquires an additional transverse momentum $\vqt$ from the medium.
The gluon's final transverse momentum with respect to the jet axis becomes $\vkt=\vlt+\vqt$.
Under the eikonal approximation for the parent parton, the $\vqt$ from the medium does not change the parent parton's momentum.
This separation between intrinsic emission momentum $\vlt$ and accumulated medium kick $\vqt$ is conceptually important.
The former is determined by the radiation dynamics, while the latter encodes post-emission rescattering.

For reconstructed jets, energy loss is not identical to partonic energy loss. Radiation that remains inside the jet cone is clustered back into the jet and does not reduce the reconstructed jet energy.
Therefore, only radiation transported outside the jet cone contributes to the measurable jet energy loss.
In the small-angle approximation, the angular separation between the emitted gluon and the jet axis can be related to its transverse momentum. 
Defining the kinematic boundary~\cite{Chang:2017gkt} $k_c\equiv 2z(1-z)E\tan\frac{R}{2}$, the condition for a gluon to lie outside the jet cone of radius $R$ can be expressed through the jet algorithm constraint $\Theta(R,\lt,\qt)=\Theta(\kt-k_c)$ where $\kt=|\vkt|$.
Radiation satisfying $\kt<k_c$ remains inside the cone and does not contribute to net energy loss.
The radiation spectrum alone determines how much energy is emitted, the jet algorithm determines how much of that energy is recovered.
The competition between intrinsic radiation scale $\lt$, accumulated diffusion scale $\qt$, and geometric constraint $k_c$ governs the sensitivity of jet observables to medium effects.

The fractional radiative energy loss of parton $a$ is therefore
\begin{align}
\label{eq:jet_EL}
\frac{\Delta E_a}{E}
&=\int\dd z\int\dd\lt^2\int\dd\tau~\left(z\frac{\dd N_a}{\dd z\dd\lt^2\dd\tau}\right)\notag\\
&\times\int\dd^2\vqt~F\left(\qt\right)\Theta\left(\kt-k_c\right).
\end{align}
The expression explicitly displays the physical mechanism of the medium-modified HT radiation kernel, transverse diffusion through Gaussian broadening, and geometric constraint by a jet cone boundary.

We note that the integral of the transverse kick can be carried out analytically, leading to a compact formulation based on Rician distribution in radial signal processing~\cite{Rice}. 
The transverse momentum kick appears only in the convolution integral
\begin{align}
\mathcal{I}(\lt,R_c)
&=\int\dd^2\vqt~F(\qt)\Theta(\kt-k_c)\notag\\
&=\int_{\kt>R_c}\dd^2\vkt~F(\vkt-\vlt).
\end{align}
Carrying out the angular integration using the Bessel identity and introducing the dimensionless variables
\begin{equation}
a^2\equiv\frac{2\lt^2}{\langle\qt^2\rangle}\quad,\quad 
b^2\equiv\frac{2k_c^2}{\langle\qt^2\rangle}\quad,\quad
x^2\equiv\frac{2\kt^2}{\langle\qt^2\rangle}\quad,
\end{equation}
the integral reduces to a one-dimensional radial integral and can be expressed compactly as
\begin{equation}
\label{eq:qt-int}
\mathcal{I}(\lt,k_c)=Q_1(a,b)~,
\end{equation}
where $Q_1(a,b)$ is the Marcum $Q$-function of order one.
For a given fixed gluon emission, $\lt$ can be treated as a radial signal and $\qt$ the noise in 2D, then $\kt$ follows a Rician distribution.
In this context, $Q_1(a,b)$ represents the probability that the radiated gluon and medium kick have transverse momentum $\kt$ outside the jet boundary $k_c$.
It also admits the convergent series expansion
\begin{equation}
Q_1(a,b)=\exp\left(-\frac{a^2+b^2}{2}\right)\sum_{n=0}^{\infty}\left(\frac{a}{b}\right)^nI_n(ab),
\end{equation}
with $I_n$ is the modified Bessel function.
For numerical stability, we can rewrite the above function as
\begin{equation}
Q_1(a,b)=\exp\left[-\frac{(a-b)^2}{2}\right]\sum_{n=0}^{\infty}\left(\frac{a}{b}\right)^n\Tilde{I}_n(ab),
\end{equation}
where $\Tilde{I}_n(z)\equiv e^{-|z|}I_n(z)$.
In the phenomenologically relevant regime where transverse diffusion is moderate and emissions satisfy $\kt\sim\lt<k_c$, one typically has $a<b$.
In this case, the series converges rapidly and can be approximated by retaining the leading terms.
In addition, in the no-diffusion limit $\langle\qt^2\rangle\to0$, $\mathcal{I}=\Theta(\lt-k_c)$, and in the collinear radiation $\lt\ll\langle\qt^2\rangle$ case $\mathcal{I}=e^{-b^2/2}$.

Substituting the result of the analytic integration into the energy loss equation~\autoref{eq:jet_EL} yields the compact expression
\begin{equation}
\frac{\Delta E_a}{E}=\int\dd z\dd\lt^2\dd\tau~\left(z\frac{\dd N_a}{\dd z\dd\lt^2\dd\tau}\right)Q_1(a,b).
\end{equation}
This form makes the physics transparent:
The HT kernel~\autoref{eq:ht-ker} governs medium-induced gluon production;
The hydrodynamic evolution enters exclusively through $\Hat{q}(\tau)$~\autoref{eq:qhat-local};
The factor $Q_1(a,b)$~\autoref{eq:qt-int} represents the probability that the emitted gluon falls outside the jet cone after transverse diffusion.
Thus, jet quenching arises from the interplay between radiation dynamics, momentum diffusion, and the geometric constraint imposed by jet reconstruction.

Energy loss depends strongly on the color representation of the propagating parton. Since $\Hat{q}_a\propto C_a$ and $\Delta E_g\approx\frac{C_A}{C_F}\Delta E_q$, gluons experience stronger broadening and enhanced radiation relative to quarks. Consequently, gluon-initiated jets are expected to lose more energy.
At leading order, the flavor of the initiating parton is uniquely determined by the hard process.
At NLO however, a reconstructed jet may originate from multiple final-state partons, making flavor assignment ambiguous.
To preserve the correct color weighting while remaining consistent with NLO kinematics, we compute the LO quark and gluon fractions $f_q(\pt)$ and $f_g(\pt)$ as functions of the initial jet transverse momentum in $\pp$ collisions.
These fractions are then used probabilistically to assign the effective jet flavor in the energy loss calculation~\cite{Han:2025ukx}.
The effective jet energy loss is therefore written as
\begin{equation}
\Delta E(\pt)=f_q(\pt)\Delta E_q+f_g(\pt)\Delta E_g,
\end{equation}
where $\Delta E_q$ and $\Delta E_g$ represent the energy loss of quark and gluon jets, respectively, which are evaluated using the corresponding color factor in $\Hat{q}_a$.


\subsection{Medium resolution and decoherence}
\label{subsec:decoh}

The medium-modified measurement function $\Tilde{\mathcal{S}}_n$ introduced in~\autoref{eq:aa-xsec} assumes that a reconstructed jet interacts with the medium as a single coherent color charge.
This approximation is valid when the medium cannot resolve the internal structure of the jet.
However, jets produced in high energy collisions generally contain multiple partons.
As a result, the interaction between the jet and the medium depends on how much of the jet's internal substructure the medium can resolve~\cite{Casalderrey-Solana:2012evi}.

Consider first two partons $i$, $j$ with transverse momenta $\pt[i]$ and $\pt[j]$, and let $d_{ij}$ be the distance measure between the two partons.
The medium can resolve both partons when their distance measure $d_{ij}$ is greater than a resolution parameter $\dc$, and their total energy loss can be written as $\Delta E(\pt[i])+\Delta E(\pt[j])$ if $d_{ij}>\dc$.
In contrast, when the medium cannot resolve the two partons $d_{ij}<\dc$, they can be viewed as a coherent cluster, and their total energy loss is $\Delta E(\pt[ij])$.
Then $\dc$ is a local characteristic parameter that reflects the resolution power of the medium.

We can then generalize this procedure to combine partons into coherent clusters for $n$ partons.
Given a set of $n$ final-state partons $\{p_1,\cdots,p_n\}$ and a medium resolution parameter $\dc$ (critical distance), the sequential coherent clustering procedure is defined as:
\begin{enumerate}[itemsep=0pt, parsep=0pt]
\item Identify the pair $i$, $j$ in the parton list with the smallest distance measure $\dij$;
\item If $\dij<\dc$, merge the pair into a single pseudo-particle $k$ and remove $i$ and $j$ from the list;
\item Repeat the above two steps until all remaining pairs satisfy the decoherent condition $\dij \geq \dc$.
\end{enumerate}
The resulting set of momenta $\{P_1,\ldots,P_m\}$ denote the resolvable coherent clusters, which act as independent color sources in the energy loss process.
For fully coherent limit ($\dc \to \infty$), all partons are merged into a single emitter.
For fully decoherent limit ($\dc \to 0$), each parton loses energy independently.
In a Monte Carlo simulation, this sequential coherent clustering procedure has to be applied to all particles at each step of the evolution to determine which cluster of particles act as a single color source in order to calculate its instantaneous energy loss.

In practice, $\dij$ can be defined in various forms.
One definition, following Ref.~\cite{Casalderrey-Solana:2012evi}, defined the pairwise distance as the pairwise angular distance $\dij \equiv \thetaij$.
Then the medium resolution $\dc=\thetac$ is the critical angle.
Note that this definition of $\dij$ only takes into account pairwise angular distance but not momenta of the parton pair. 
In this study, we define the pairwise distance as the pair invariant mass to determine coherence:
\begin{equation}
d_{ij}^2 \equiv \Qij^2 = (p_i+p_j)^2 \approx \pt[i]\pt[j] \thetaij^2,
\end{equation}
where the medium resolution parameter $\dc=\Qc$ (critical scale) is now a dimensionful quantity containing both angular and momentum information, and is related to the critical angle via $\Qc^2\approx\pt[i]\pt[j] \thetac^2$.
Such that for a pair of partons with $\Qij<\Qc$, both partons lose energy as a whole, while for $\Qij\geq\Qc$ both partons lose energy independently.

The critical scale $\Qc$ is in principle dependent on the transport properties of the medium and the path that the parton travels, thus it could take different values for each step in the coherent cluster procedure.
To model fluctuations of the effective resolution, two distributions of the critical scale $P(\Qc)$ can be parameterized as follows:
\begin{equation}
P(\Qc) =
\begin{cases}
\delta (\Qc - \Qmed ) \\
\frac{2\Qmed^2}{\Qc^3} \exp\left[-\left(\frac{\Qmed}{\Qc} \right)^2\right] 
\end{cases},
\label{eq:P_Qc}
\end{equation}
where $\Qmed$ is a fixed effective medium resolution scale, which can be determined by fitting with experimental data.
The first distribution is simply setting the critical scale to the effective medium resolution scale, assuming the magnitude of the medium resolution is fixed.
The second distribution assumes that the magnitude of the critical scale follows an inverse Rayleigh distribution, with $\Qmed$ its instantaneous magnitude.

This allows us to define a coherence intensity function
\begin{equation}
\Ico(\Qij) = \int_{\Qij}^{\infty}\dd\Qc~P(\Qc),
\end{equation}
where it represents the probability that a pair of partons with invariant mass $\Qij$ being unable to be resolved by the medium with an effective resolution scale $\Qmed$.
The two parameterizations above gives the following coherence intensity:
\begin{equation}
\label{eq:Ico}
\Ico(\Qij) =
\begin{cases}
\Theta(\Qmed - \Qij),    &\text{step form}\\
 1-\exp\left[-\left(\frac{\Qmed}{\Qij} \right)^2\right], &\text{smooth form}
\end{cases}.
\end{equation}

Consider the ``step form'' case, where $\Qc=\Qmed$ is fixed for the collision.
We can apply the coherent clustering procedure to all particles to determine the number of coherent cluster that the medium can ``see'' in order to calculate the energy loss of the cluster.
In our calculation, we distribute the energy loss of the coherent cluster to each parton proportional to its energy fraction, i.e. $\mathcal{Q}(p)=p-\frac{p}{P}\Delta E(P)$.
Thus, $\mathcal{Q}(\pt[i])$ is the transverse momentum of the final observed particle, which then goes into the jet clustering algorithm.
It is important to note that the coherent clustering procedure and jet algorithm are independent, and the implementation of coherent energy loss allow particles that are not within a jet, to have influence over the energy loss of particles inside the jet.
For example, only particle $i$ in a coherent pair $ij$ may be clustered into the jet but not particle $j$, then the energy loss of particle $i$ depends on the momentum of particle $j$, i.e. $\mathcal{Q}(\pt[i])=\pt[i]-\frac{\pt[i]}{\pt[ij]}\Delta E(\pt[ij])$.

For the ``smooth form'', both coherent and decoherent case of energy loss have to be considered for each step in the coherent clustering procedure, where the probability of each case is calculated by the coherent intensity.
For $n$-particles, $n$ cases have to be considered.
Take 3-particles for example, and let $d_{ij}$ be the smallest distance, then the probability of the three cases are: $\Ico(Q_{ij})\Ico(Q_{(ij)k})$, $\Ico(Q_{ij})\left[1-\Ico(Q_{(ij)k})\right]$, and $\left[1-\Ico(Q_{ij})\right]$.
Because this form of $\Ico$ is non-zero for large $\Qij$, coherent energy loss of two particles is suppressed but not identically zero even if they are separated by a large distance $\dij$.

The invariant-mass criterion ensures that soft and collinear partons are clustered into the same coherent structure in the corresponding limits. 
This construction is designed to preserve the infrared and collinear behavior of the measurement function. 
For the step-form, the procedure is consider IRC safe at the algorithmic level as long as the critical scale is sufficiently large $\Qc^2\gg\Lambda^2$.
The singular behaviors $\Qij\to0$ are entirely contained in the coherent regime, and the measurement function $\Tilde{\mathcal{S}}_n$ remains finite and well-defined over the full phase space, preserving convergence at the cross-section level.
A complete demonstration of IRC safety, however, requires considering the corresponding limits of the full medium-modified measurement function.

Physically, $\Qc$ encodes the characteristic scale at which the medium transitions from coherent to resolving behavior.
One may expect that $\Qc$ is related to medium properties such as the local temperature, density, and path length.
In particular, dimensional analysis and insights from in-medium antenna studies suggest that it may be connected to the transport coefficient through relations of the form $\Qc^2\sim\pt[2]/(\Hat{q}L^3)$.
Within this interpretation, $\Qc$ characterizes the resolving power of the QGP to the jet, where small values correspond to a highly resolving medium in which substructures lose energy independently, while large values correspond to a more coherent regime.
The following section is dedicated to the determination of the effective medium resolution $\Qmed$ by comparison with the experimental data.


\section{Numerical Analysis}
\label{sec:result}
In this section, we begin by analyzing the coherent clustering procedure on \texttt{PYTHIA} generated data in $\pp$ collisions.
To investigate the suppression of full jets, we calculate the $\RAA$ of inclusive jets as a function of $\pTjet$ in $\PbPb$ collisions at $\sNN=5.02$ TeV, and extract the value of $\Qmed$.
The reclustered jet $\RAA$ in $\PbPb$ collisions are calculated to investigate the influence of decoherent energy loss on jets with both single subjet and multiple subjets.
To further explore the impact of decoherent energy loss on the jet substructure, we calculate the suppression of both soft-drop $\rg$ and jet mass in the same collisions system.


\subsection{Decoherent clustering}
\label{subsec:decoh_clustering}

\begin{figure}[htbp]
\centering
\includegraphics[width=\linewidth]{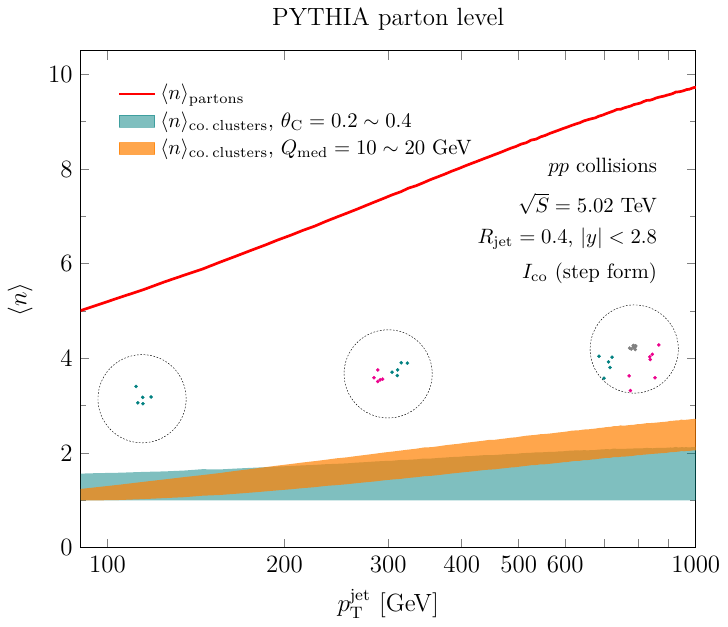}
\caption{
The above plot shows the mean multiplicity within a \texttt{PYTHIA} simulated jet of $R=0.4$ using anti-$k_\mathrm{t}$ algorithm in $\pp$ collisions at $\sqrt{S}=5.02$ TeV.
The average number of partons (red solid), coherent clusters using angle (teal band), and coherent clusters using invariant mass (orange band) are plotted as a function of the jet $\pTjet$.
Three selected event with 1, 2, and 3 coherent clusters are also illustrated as reference.
}
\label{fig:pythia}
\end{figure}

Multi-particle final states from parton showers simulated in a \texttt{PYTHIA}-generated event provides a testing ground for the sequential coherent clustering procedure introduced in~\autoref{subsec:decoh}.
We begin by setting \texttt{PYTHIA}~\cite{Bierlich:2022pfr} with \texttt{A21} tune matched with the ATLAS~\cite{ATLAS:2019rqw}, with hadronization turned off to focus only on the final parton shower effect.
Jets are then reconstructed with the anti-$k_t$ algorithm~\cite{Cacciari:2008gp} using the \texttt{FastJet} package~\cite{Cacciari:2011ma} with $R_\mathrm{jet}=0.4$ in $pp$ collisions at $\sqrt{S}=5.02$ TeV.
The mean parton multiplicity (red solid line) is shown in ~\autoref{fig:pythia} as a function of the jet $\pTjet$.

We first applied the coherent clustering procedure using pairwise angular separation as the distance measure $\dij=\thetaij$, and plotted the mean multiplicity of coherent clusters (teal band) by setting the critical angle to be $\thetac=0.2\sim0.4$.
One can see that at low $\pTjet$, the mean multiplicity is one, consistent with the coherent energy loss picture.
At high $\pTjet$ however, the mean multiplicity increased to 2, suggesting two coherent color sources within a jet that loses energy independently.

Then, we applied the coherent clustering procedure with the pair invariant mass as the distance measure $\dij=\Qij$.
By fixing the critical scale with an effective medium resolution parameter $\Qc=\Qmed$ (step form), we plotted the mean multiplicity of coherent clusters (orange band) with the effective scale set to $\Qmed=10-20$ GeV.
Similar to the above procedure, the mean multiplicity grows as a function of the jet $\pTjet$.
However, the mean multiplicity using pair invariant mass grows faster compared to the above procedure, and a depletion of single coherent clusters within a jet is also seen at high $\pTjet$.
At low $\pTjet$ ($<200$ GeV), jets have on average one single coherent cluster, consistent with the coherent energy loss picture.
At intermediate $\pTjet$ ($300\sim800$ GeV), jets are dominated by two-prong structures, and one has effectively twice the energy loss.
At higher $\pTjet$ ($>900$ GeV), jets have three or more coherent cores that loses energy independently.

Both procedures showed that low $\pTjet$ jets are predominantly single coherent cluster jets, and high $\pTjet$ jet exhibit an increasing probability of containing multiple resolved substructures.
This behavior indicates that decoherent mechanism of jet energy loss are indispensable in describing nuclear modifications and jet-medium interactions at high $\pTjet$, and this description will lead to stronger energy loss effect relative to the coherent limit.

Finally, three jets were selected at 0.1, 0.5, and 1.0 TeV with 1, 2, and 3 coherent clusters to highlight the spatial distribution of particles that get clustered within a jet.
This is shown as three circles in~\autoref{fig:pythia} where same colored dots representing partons belong to the same coherent cluster.
Although the procedure is oblivious to the spatial distribution of the particles and their radiation order, it automatically cluster particles that are spatially together.
This might be due to the fact that angular ordering is implemented in the showering mechanism, and the coherent clusters are approximately those initial hard radiations in the shower.

\subsection{Inclusive Jet \texorpdfstring{$R_{AA}$}{RAA}}
\label{subsec:inclusive_jet}

Numerous observables in the heavy-ion collision experiments provide clear evidence of jet energy loss.
To characterize inclusive jet energy loss in $\AA$ collisions, we compute the nuclear modification factor $\RAA$ of single inclusive jets, incorporating both medium-induced energy loss and the decoherence effects introduced in~\autoref{sec:theory}.
The observable is defined as
\begin{equation}
R_\AA (\pt) = \frac{1}{\int \, T_\AA(\vbp) \, \dd^2 \vbp} \,
\frac{\dd \sigma^\AA_\mathrm{jet}/\dd \pTjet \dd y}{\dd \sigma^\pp_\mathrm{jet}/\dd \pTjet \dd y} ,
\end{equation}
where $\dd \sigma^\pp_\mathrm{jet}/\dd \pTjet$ and $\dd \sigma^\AA_\mathrm{jet}/\dd \pTjet$ denote the jet $\pTjet$ spectra in $\pp$ and $\AA$ collisions, respectively.
The integral of $T_{AA}$ over the centrality class normalizes the $\AA$ cross-section over the number of binary collisions.
The $\pp$ baseline is computed at $2\to2$ NLO accuracy, while the heavy-ion result includes both nPDF effects~\cite{Eskola:2021nhw, Pumplin:2002vw, Giele:2001mr} and jet energy loss.

\begin{figure}[htbp]
\centering
\includegraphics[width=\linewidth]{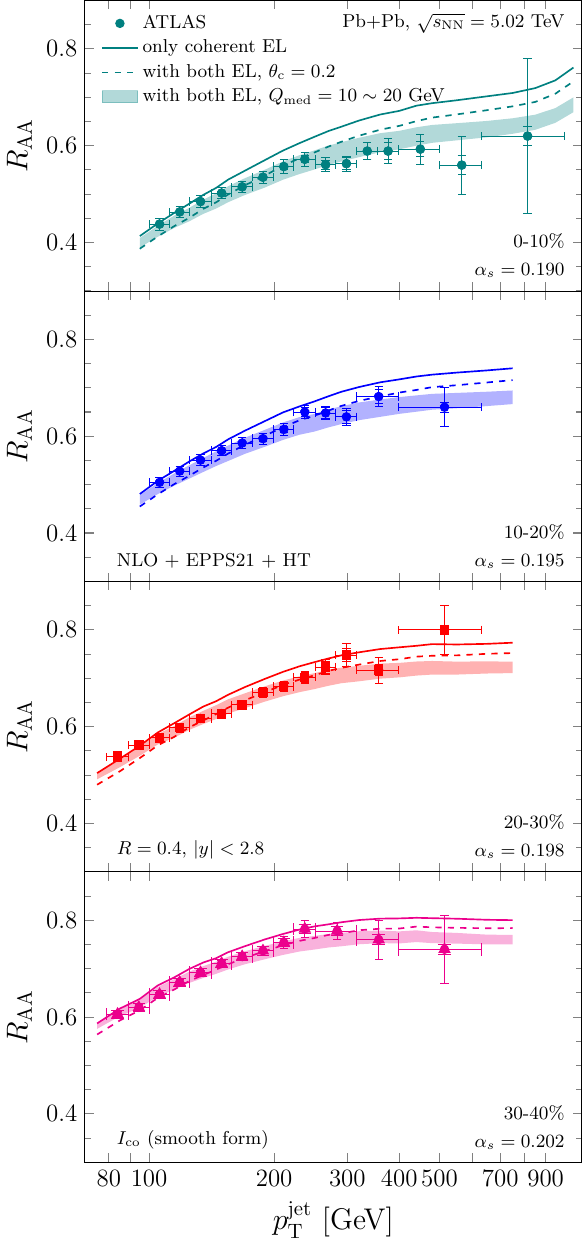}
\caption{
Nuclear modification factor $\RAA$ of single inclusive jets as a function of the jet $\pTjet$ in $\sNN=5.02$ TeV $\PbPb$ collisions for four different centralities. 
The theoretical calculation of the coherent energy loss (solid), both coherent and decoherent under angular separation criterion (dashed), and under invariant mass criterion (band), respectively, are plotted against the ATLAS experimental data~\cite{ATLAS:2018gwx}.}
\label{fig:RAA_inclusive_jet}
\end{figure}

To observe the decoherent energy loss in $\PbPb$ collisions at $\sNN=5.02$ TeV, we first calculate the nuclear modification factor $\RAA$ of inclusive jets as a function of jet $\pTjet$ shown in~\autoref{fig:RAA_inclusive_jet}, where the inclusive jets with $\Rjet=0.4$ and rapidity cut $|y|<2.8$ are compared with experimental data in four centralities (0-10\%, 10-20\%, 20-30\% and 30-40\%) in $\PbPb$ collisions measured by the ATLAS collaboration~\cite{ATLAS:2018gwx}.

The coherent baseline is set by assuming that low $\pTjet$ jets are dominated by coherent clusters and the calculation with coherent energy loss can describe the experimental data with the lowest $\pTjet$.
This allows us to fixed the strong coupling $\as$ that enters in the Higher-Twist energy loss formalism at different centralities similar to~\cite{Han:2025ukx}.
Setting the coherent energy loss (solid line) to go through the data with the lowest $\pTjet$, we see that the effective coupling responsible for describing the medium induced radiation increases toward more peripheral collisions~\cite{Xie:2022ght, Xie:2022fak}, and the coherent-only calculation gives larger $\RAA$, thereby underestimating the suppression of the experimental data at high $\pTjet$ especially in the most central collisions.
This indicates that high $\pTjet$ jets likely contain multiple substructures that can be resolved by the medium, and their energy loss strengthens the quenching effect and therefore, the coherent approximation will fail in this regime.

To account for this effect, we include the decoherence mechanism through the medium modified measurement function, and the transition between coherent and decoherent energy loss is governed by the coherence condition $\dij<\dc$ in~\autoref{subsec:decoh}.
In~\autoref{fig:RAA_inclusive_jet}, the $\RAA$ with both coherent and decoherent energy loss is plotted for both the angular criterion $\thetaij$ and the invariant mass criterion $\Qij$.
We note that using the angular criterion for coherence and setting $\thetac=0.2$ to a fixed value only result in an overall shift of the $\RAA$ spectrum.
But when we use the pair invariant mass as criterion for coherence, we see a stronger suppression of the $\RAA$ at high $\pTjet$.
This is because high $\pTjet$ jets are more likely to have splittings with large virtuality, and resulting partons lose energy independently.
By comparison with the ATLAS experimental data, we extracted the range of the effective medium resolution scale to be $\Qmed=10-20$ GeV.
We also note that in our numerical calculation, both step form and smooth form of the coherent intensity $\Ico(\Qij)$ give similar result due to the insensitivity of the $\RAA$ observable.

This result demonstrates that medium resolution effects are essential for a quantitative description of inclusive jets suppression especially for jets with high $\pTjet$.
In particular, it suggests that the QGP has a characteristic dimensionful scale that characterizes its resolution power and the coherent clusters within jets are only resolved when their invariant mass are above this characteristic scale, which leads to an effective increase in the number of independent color sources participating in energy loss.

\subsection{Reclustered Jet \texorpdfstring{$R_{AA}$}{RAA}}
\label{subsec:reclustered_jet}

To further probe the role of coherent clusters in medium induced energy loss, we consider observables that explicitly classify jets according to their internal structure.
Such measurements provide a more differential test to the decoherent mechanism.
The reclustered jets measured by the ATLAS~\cite{ATLAS:2023hso} experiment maps naturally to this decoherent picture, where particles were first clustered by the anti-$k_t$ algorithm to producing $\Rsubjet=0.2,~\pt[\rm subjet]>35$ GeV subjets, then the algorithm with $\Rjet=1.0$ was applied again to produce jets.
Jets were then classified into two different categories: jets containing only one subjet, and jets containing two or more subjets.

\begin{figure}[htbp]
\centering
\includegraphics[width=\linewidth]{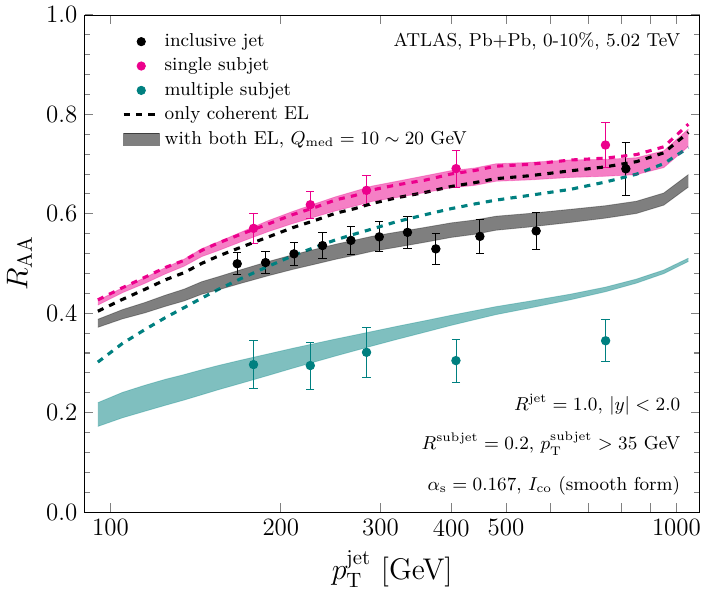}
\caption{The nuclear modification factors $\RAA$ in 0-10\% central $\PbPb$ collisions as a function of $\pTjet$ for reclustered jets. $\Rjet=1.0$ reclustered jets with a single subjet (magenta), multiple subjets (teal), and inclusive (black), measured by the ATLAS~\cite{ATLAS:2023hso} experiment with subjets $\Rsubjet=0.2$ and $\pt[\rm subjet]>35$ GeV.
With only coherent (dashed lines), and with both coherent and decoherent (band) energy loss calculations were also plotted.}
\label{fig:RAA_reclustered_jet}
\end{figure}

For our calculation, we set the coherent baseline by calculating the $\RAA$ spectrum with $\Rjet=0.2$ jets in $\PbPb$ collisions, where the result is matched to the ATLAS single subjet data at the smallest $\pTjet$ and set the strength of the medium-induced radiation to $\as=0.167$ similar to the previous subsection~\autoref{subsec:inclusive_jet}.
The reason for this is because we assume that coherent quenching are dominant at low $\pTjet$, and the $\RAA$ spectrum of $\Rjet=0.2$ is similar to $\Rjet=1.0$ jets that contain only one $\Rsubjet=0.2$ subjet in this regime.
This is shown as the magenta dashed line in~\autoref{fig:RAA_reclustered_jet}, and we see a very good agreement between the ATLAS data and the coherent energy loss calculation.
The decrease in the strong coupling from 0.190 to 0.167 in comparison to the previous subsection suggests that $\Rjet=0.2$ jets lose more energy than $\Rjet=0.4$ jets in the coherent limit due to more radiation falling outside the jet cone.
However, the coherent energy loss calculation fails to describe $\RAA$ for inclusive jets and jets containing multiple subjets, and this is evident in the plot where coherent calculation (black and teal dashed lines) underestimated the strength of energy loss.

To account for the quenching effect of multiple subjets, which is similar to the decoherent energy loss effect, We employed our earlier calculation under the decoherent framework following the ATLAS reclustering, and use the effective medium resolution parameter extracted from the previous subsection $\Qmed=10-20$ GeV to determine whether partons gets quenched individually or coherently.
The result is plotted as the bands in~\autoref{fig:RAA_reclustered_jet}, where we see a very good agreement between the ATLAS experimental data and the calculation with both coherent and decoherent energy loss, in comparison to only the coherent energy loss especially for inclusive jet and jets with multiple subjets.

We see that the calculation with both coherent and decoherent energy loss for single subjet is almost identical to the coherent calculation because $\Rjet=0.2$ jets can be thought of as a coherent clustered in a medium with resolution $\Qmed=10-20$ GeV.
However, jets with multiple subjets can only be described by the energy loss framework with both coherence and decoherence because the medium can resolve the substructures of the jet and they lose energy independently.
We note that the effects of nPDF becomes important at high $\pTjet$, where particle yields will experience a suppression due to the EMC effect in the $<800$ GeV region, and an enhancement in the $>800$ GeV region.
Because the calculation of the jet $\pTjet$ spectrum at NLO corresponds to $2\to3$ process, which means at most 2 partons in a jet, our decoherent energy loss calculation underestimated the energy loss effect for jets with multiple subjets at very high $\pTjet$.

\subsection{Jet \texorpdfstring{$\rg$}{rg}}
\label{subsec:jet_rg}

Jet substructure observables provide a direct way to investigate how the medium resolves the internal structure of a jet and how this resolution affects its energy loss.
In particular, the groomed jet radius $\rg$ characterizes the angular separation of the two subjets associated with the first hard splitting of the jet.
We employ the soft-drop grooming procedure~\cite{Larkoski:2014wba}, with $\zcut=0.2$ and $\beta=0$, to identify this splitting.
The final value of $\rg$ is defined as the angular separation $\Delta R_{12}$ between the two subjets that satisfy the soft-drop condition.
If no splitting passes the condition, the jet is regarded as having no resolvable hard two-prong structure, and $\rg$ is set to zero.

\begin{figure}[htbp]
\centering
\includegraphics[width=\linewidth]{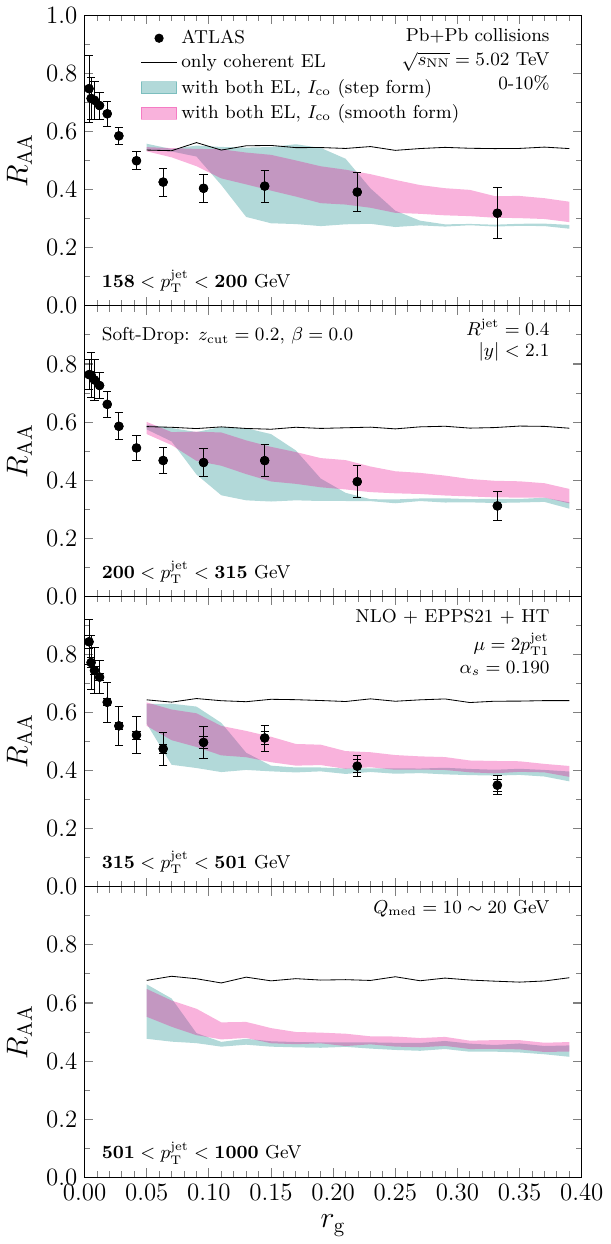}
\caption{
Nuclear modification factor $\RAA$ as a function of the soft-drop groomed radius $\rg$ in 0-10\% central $\PbPb$ collisions at $\sNN=5.02$ TeV, for different jet $\pTjet$ intervals.
The black lines represent predictions with only coherent energy loss, while the teal and magenta bands include both coherent and decoherent energy loss with the step and smooth forms of $\Ico$, respectively, using $\Qmed=10-20$ GeV.
ATLAS measurements~\cite{ATLAS:2022vii} are shown for comparison.
}
\label{fig:RAA_jet_rg}
\end{figure}

\autoref{fig:RAA_jet_rg} presents the nuclear modification factor $\RAA$ as a function of $\rg$ in 0-10\% central $\PbPb$ collisions at $\sNN=5.02$ TeV, for four different jet $\pTjet$ intervals.
The black solid curves show the results obtained with coherent energy loss only.
Within this scenario, the energy loss is determined by the total momentum of the coherent jet and is insensitive to its internal substructure within our model.
Consequently, $\RAA$ exhibits little dependence on $\rg$ within each $\pTjet$ interval.
This behavior does not reproduce the observed variation of the nuclear modification factor with the groomed jet radius.

The inclusion of decoherent energy loss leads to a qualitatively different behavior.
When the medium resolves the coherent clusters, they can lose energy independently, and the resulting suppression becomes sensitive to the angular structure of the jet. 
The predictions incorporating both coherent and decoherent energy loss, shown by the teal and magenta bands, therefore exhibit a nontrivial dependence on $\rg$ and provide a substantially improved description of the ATLAS measurements~\cite{ATLAS:2022vii}. 
This comparison demonstrates that coherent energy loss alone is insufficient to describe the substructure dependence of jet suppression, and that medium-induced decoherence is essential for capturing the observed modification.

We further compare the two parameterizations of the coherent intensity $\Ico(\Qij)$. 
The step form produces a relatively sharp variation in $\RAA$, reflecting the use of a fixed resolution threshold. 
In contrast, the smooth form leads to a more gradual dependence on $\rg$ and gives a closer description of the experimental data. 
We interpret this smooth parametrization phenomenologically: it accounts for fluctuations in the effective resolution scale of the medium, rather than representing a first-principles calculation of those fluctuations. 
The comparison thus suggests that a description incorporating a distribution of effective medium-resolution scales is more appropriate than one based on a sharp resolution boundary.

\subsection{Jet invariant mass}
\label{subsec:jet_mass}

The jet invariant mass provides another probe of the internal momentum and angular structure of a jet~\cite{Duan:2023gmp}. 
It is defined as $m=\sqrt{E^2-p^2}$ where $E$ and $p$ are the total energy and three-momentum of the reconstructed jet, respectively. 
In the following, we consider the dimensionless observable $m/\pTjet$, following the definition adopted in the experimental measurement~\cite{ATLAS:2018jsv}. 
Its dependence on the jet invariant mass provides complementary information on how the medium-induced energy loss is affected by the internal structure of the jet.

\begin{figure}[htbp]
\centering
\includegraphics[width=\linewidth]{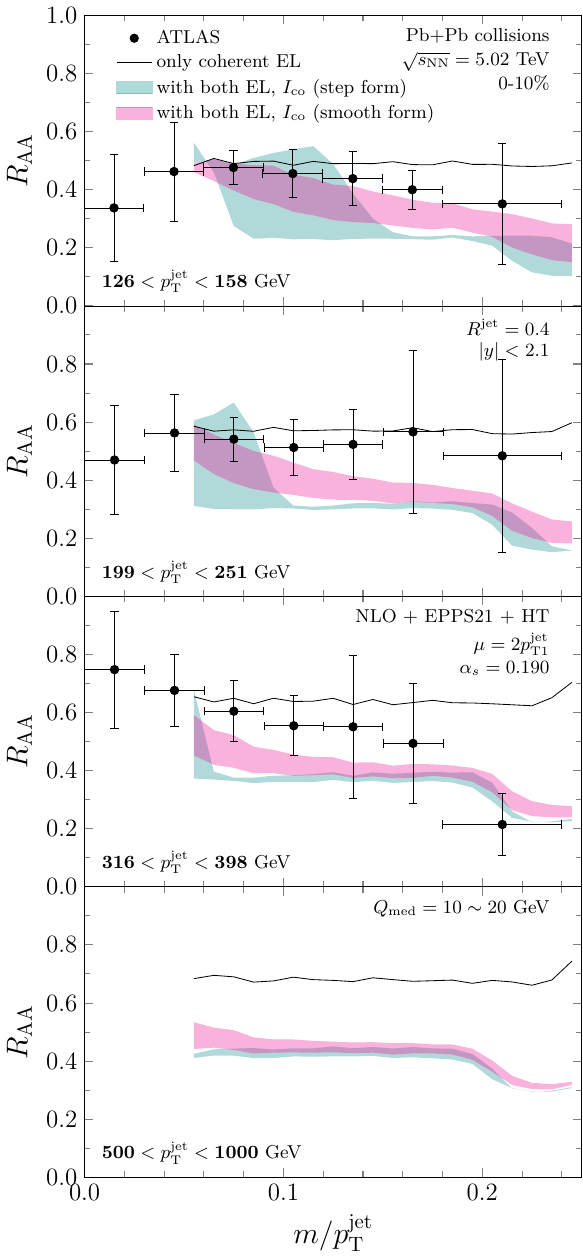}
\caption{
Nuclear modification factor $\RAA$ as a function of the normalized jet invariant mass $m/\pTjet$ in 0-10\% central $\PbPb$ collisions at $\sNN=5.02$ TeV, for different jet $\pTjet$ intervals.
The black lines represent predictions with only coherent energy loss, while the teal and magenta bands include both coherent and decoherent energy loss with the step and smooth forms of $\Ico$, respectively, using $\Qmed=10-20$ GeV.
ATLAS measurements~\cite{ATLAS:2018jsv} are shown for comparison.
}
\label{fig:RAA_jet_mass}
\end{figure}

\autoref{fig:RAA_jet_mass} shows the nuclear modification factor $\RAA$ as a function of the normalized jet mass $m/\pTjet$ in 0–10\% central $\PbPb$ collisions at $\sNN=5.02$ TeV, for several intervals of jet transverse momentum. 
The black solid curves correspond to the calculation including only coherent energy loss. 
As in the case of $\rg$, this contribution is largely insensitive to the internal structure of the jet and consequently gives an approximately constant $\RAA$ over the considered range of $m/\pTjet$. 
The coherent-only calculation therefore fails to reproduce the observed dependence of jet suppression on the invariant mass.

When decoherent energy loss is included, the predictions exhibit a stronger suppression and a clear dependence on jet mass at large $m/\pTjet$. 
This behavior arises because the energy loss of a jet is no longer determined solely by its total momentum: when the medium resolves its internal coherent clusters, the individual constituents can lose energy independently. 
The resulting suppression is therefore sensitive to the partonic configuration and branching structure inside the jet. 
As shown in~\autoref{fig:RAA_jet_mass}, the calculations incorporating both coherent and decoherent energy loss provide a substantially better description of the ATLAS data~\cite{ATLAS:2018jsv} than the coherent-only result. 
This comparison further supports the importance of medium resolution and decoherent energy loss in describing jet-substructure modifications in heavy-ion collisions.

The difference between the two forms of $\Ico(\Qij)$ is also visible in the mass dependence. 
The step parametrization gives a relatively sharp transition in the predicted $\RAA$, whereas the smooth parametrization produces a more gradual variation with $m/\pTjet$ and is in better agreement with the experimental measurements. 
As in the $\rg$ analysis, the smooth form is a phenomenological parametrization intended to account for fluctuations in the effective medium-resolution scale. 
It should not be interpreted as a microscopic derivation of the fluctuations themselves. 
The consistent improvement obtained with this parametrization in both substructure observables indicates that the data favors a description in which the effective resolution scale is not represented by a single sharp threshold within this model.

Finally, we note that the perturbative accuracy of the calculation places a limitation on the partonic multiplicity available for the substructure analysis. 
Since jet $\rg$ and the jet invariant mass are two-particle observables which require at least three particles in the final state to obtain a nontrivial distribution, we employ the $2\to3$ NLO calculation, which includes up to $2\to4$ real-emission contribution and allows at most three final-state partons to be clustered into a jet. 
Contributions from configurations with higher partonic multiplicities are therefore not included. 
This limitation should be kept in mind when interpreting the quantitative predictions, particularly in regions where jets with more complicated internal structures may contribute significantly.

\section{Conclusion}
\label{sec:conclusion}

In this work, we develop a resolution-based framework to investigate coherent and decoherent jet energy loss in high-energy heavy-ion collisions.
We establish the $\pp$ baseline using NLO pQCD calculations implemented in \texttt{NLOJet++}, with $2\to2$ NLO processes for inclusive and reclustered jets, and $2\to3$ NLO processes for the groomed-radius and jet-mass observables.
For $\AA$ collisions, we incorporate nuclear PDFs, Glauber geometry, and medium-induced jet energy loss within the Higher-Twist formalism.

To account for the finite resolution of the QGP, we introduce a sequential coherent clustering procedure based on the pairwise invariant mass $\Qij$ of final-state partons.
Partons with invariant masses below the medium resolution scale are combined into coherent clusters, which subsequently lose energy as independent color sources when resolved by the medium.
The coherence intensity $\Ico(\Qij)$ is further introduced in step and smooth forms to account for different parameterizations of fluctuations in the medium resolution scale.
The coherent clustering procedure is validated using \texttt{PYTHIA} simulations, which demonstrate an increasing number of resolvable coherent clusters with increasing jet transverse momentum.

We investigate the nuclear modification factor of inclusive jets, reclustered jets, soft-drop $\rg$, and jet mass in $\PbPb$ collisions at $\sNN=5.02$ TeV, and compare our predictions with ATLAS measurements.
For inclusive jets, the coherent energy loss approximation provides a reasonable description at low $\pTjet$, but systematically underestimates the suppression at high $\pTjet$.
Incorporating decoherent energy loss improves the agreement with the data and an effective medium resolution scale of $\Qmed=10-20$ GeV. 
The reclustered jet measurements further demonstrate that jets containing multiple resolved subjets require decoherent energy loss, whereas single subjet jets remain predominantly in the coherent regime.

The substructure observables $\rg$ and $m/\pTjet$ provide additional constraints on the resolution mechanism.
While coherent energy loss leads to an approximately constant $\RAA$ over the measured substructure ranges, the inclusion of decoherent energy loss generates a nontrivial dependence on both observables and significantly improves the description of the experimental measurements.
Furthermore, the smooth-form coherence intensity generally reproduces the gradual suppression patterns more accurately than the step-form parametrization, suggesting that fluctuations in the medium resolution scale play an important role in jet-medium interactions.

Overall, our results demonstrate that a unified description of inclusive jet suppression and jet substructure modifications requires both coherent and decoherent energy loss mechanisms.
The resolution-based coherent clustering procedure provides a practical framework for incorporating the finite resolving power of the QGP into NLO pQCD jet calculations and offers a means of constraining medium properties through differential jet observables.

\begin{acknowledgments}
The authors would like to thank Y. Shi for helpful discussions. 
This work is supported by National Natural Science Foundation of China under Grant No.~12535010.
L.C. would also like to thank C.A. Salgado for helpful discussions.
L.C. is funded by the European Union’s Horizon Europe research and innovation program under the Marie Sk\l{}odowska-Curie grant agreement No.~101210595.
\end{acknowledgments}

\bibliographystyle{apsrev4-2}
\bibliography{reference}

\end{document}